\documentclass[12pt]{article}
\usepackage{amssymb}
\usepackage{amsmath}
\usepackage{epsfig}
\usepackage{graphicx}
\usepackage{wrapfig}

\begin{document}

\newcommand{\ppbch}{$pp \to \bar bcH^- +X$ }
\newcommand{\ppbchT}{$pp \to bcH^\pm +X$ }
\newcommand{\ppbthT}{$pp \to btH^{\pm} +X$ }
\newcommand{\qgbch}{$q\bar q/gg \to \bar bcH^-$ }

\title{Flavor changing effects on single charged Higgs boson production
associated with a bottom-charm pair at CERN Large Hadron Collider
\footnote{Supported by National Natural Science Foundation of
China.}} \vspace{3mm}

\author{\small{ Sun Hao, Ma Wen-Gan, Zhang Ren-You, Guo Lei, Han Liang and Jiang Yi}\\
{\small Department of Modern Physics, University of Science and Technology}\\
{\small of China (USTC), Hefei, Anhui 230026, P.R.China}  }

\date{}
\maketitle \vskip 12mm

\begin{abstract}
We study flavor changing effects on the \ppbchT process at the
Large Hadron Collider(LHC), which are inspired by the left-handed
up-type squark mixings in the Minimal Supersymmetric Standard
Model(MSSM). We find that the SUSY QCD radiative corrections to
$bcH^\pm$ coupling can significantly enhance the cross sections at
the tree-level by a factor about $1.5 \sim 5$ with our choice of
parameters. We conclude that the squark mixing mechanism in the
MSSM makes the \ppbchT process a new channel for discovering a
charged Higgs boson and investigating flavor changing effects.
\end{abstract}

\vskip 1cm

{\large\bf PASC: 12.60.Jv, 14.80.Cp, 14.65.Fy}

\vfill \eject

\baselineskip=0.36in

\renewcommand{\theequation}{\arabic{section}.\arabic{equation}}
\renewcommand{\thesection}{\Roman{section}}
\newcommand{\nb}{\nonumber}

\makeatletter      
\@addtoreset{equation}{section}
\makeatother       

\section{Introduction}
\par
As we know there are stringent experimental constraints against
the existence of tree-level flavor changing scalar
interactions(FCSI's) for light quarks. It leads to the suppression
of the flavor changing neutral current(FCNC) couplings at the
lowest order. This is also an important feature of the standard
model(SM)\cite{SM}. Even the one-loop flavor changing effect in
the SM is still small, due to the suppression of the
Glashow-Iliopulos-Maiani(GIM) mechanism\cite{GIM}. In extension
models beyond the SM when new non-standard particles exist in the
loops, significant contributions to flavor changing transitions
may appear. Among various new physics models, the minimal
supersymmetric extension(MSSM)\cite{MSSM} is widely considered as
the one of the most appealing extensions of the SM. The MSSM not
only can explain the existing experimental data as the SM does,
but also can be used to solve various theoretical problems in the
SM, such as the huge hierarchy problem between the electroweak
symmetry-breaking and the grand unification scales. In the MSSM
there exist two Higgs doublets to break the electroweak symmetry.
After symmetry breaking, there are five physical Higgs bosons: two
CP-even Higgs bosons($h^0, H^0$), one CP-odd boson($A^0$)and two
charged Higgs bosons($H^\pm$)\cite{Higgsbosons}. Significant
difference exists between the couplings involving Higgs boson in
the MSSM and those in the SM. In SUSY models, an important feature
is that the fermion-Higgs couplings are no longer strictly
proportional only to the corresponding mass as they are in the SM.
For example, the b-quark coupling with neutral Higgs boson $A^0$
in the MSSM becomes enhanced for large $\tan\beta=v_2/v_1$, the
ratio of the two vacuum expectation values\cite{Higgsbosons}.
Thus, different features can be presented due to the existence of
the five Higgs bosons in the MSSM which might lead different
coupling strengths, decay widths and production cross sections
compared with in the SM.

\par
The understanding of flavor sector is a major challenge for
various extension models of the SM. In the MSSM, the minimal
flavor violation is realized by the CKM-matrix\cite{CKM}. While in
the general MSSM with flavor violation, a possible flavor-mixings
between the three sfermion generations are allowed and lead to
flavor changing effects. The flavor changing effects originating
from such sfermion-mixing scenario normally cannot be generated at
the tree-level, but could show up at the one-loop level and induce
significant contributions to be observed in specific regions of
the MSSM parameters.

\par
Searching for scalar Higgs bosons is one of the major objectives
of present and future high energy experiments. In most extensions
of the SM, the mass of a charged Higgs boson($m_{H^\pm}$) is
predicted to be around the weak scale. However, the Higgs bosons
haven't been directly explored experimentally until now. At hadron
colliders, such as the Fermilab Tevatron and the CERN Large Hadron
Collider(LHC), a light charged Higgs boson can be produced from
the decay of top quark via $t\to H^+ b$, if
$m_{H^\pm}<m_t-m_b$\cite{Higgs}. Otherwise, if the charged Higgs
boson is heavier than top quark, there are three major channels to
search for charged Higgs boson: (1)charged Higgs boson pair
production\cite{HiggsH,HiggsH1,HiggsH2}; (2)associated production
of a charged Higgs boson with a W boson\cite{HiggsW};
(3)associated production of a charged Higgs boson with a top quark
$gb \to tH^-$\cite{HiggsT}; (4)single charged Higgs production
$\bar cs,\bar cb \to H^-$\cite{HiggsC}. The decay of the charged
Higgs boson has two major channels: $H^- \to \bar tb$\cite{Htb},
and $H^- \to \tau^- \bar \nu$\cite{Htn}. At the LHC, the most
promising channel to search for the charged Higgs boson in some
specific parameter space, is \ppbthT, whose QCD corrections have
been studied in Ref.\cite{wpSMQCD,wpSUSYQCD,wp}. The \ppbchT
process is another important alternative channel especially
considering the contributions from squark-guino loops with flavor
mixing structure. J.L. Diaz-Cruz, et al., analyzed SUSY radiative
corrections to the $bcH^\pm$ and $tch^0$ couplings including
squark-mixing effects, and showed that these couplings can reveal
exciting new discovery channels for the Higgs boson signals at the
Tevatron and the LHC\cite{HiggsC}. H.J. He, et al,. studied the
single charged Higgs production process at linear colliders, such
as $e^-e^+ \to b \bar c H^+$, $\tau \bar \nu H^+$ and
$\gamma\gamma \to b \bar c H^+$, $\tau \bar \nu
H^+$\cite{HiggsC1}. The flavor changing effect on the neutral
Higgs boson production associated with a bottom-strange quark pair
in the MSSM at the linear collider was studied in
Ref.\cite{eehbs}.

\par
As we know, among the three generations of fermions, the top-quark
is the heaviest one with its mass as high as the electroweak
scale. The large top-quark mass will enhance flavor changing
Yukawa coupling $bcH^\pm$ at the loop level and make the single
charged Higgs production process \ppbchT to be an important
channel for probing flavor violation and searching for charged
Higgs boson at hadron colliders. Furthermore, when the neutral
scalar($\phi^0$) and the charged scalar ($\phi^\pm$) form a SU(2)
doublet, the weak isospin symmetry connects the flavor-changing
neutral coupling(FCNC) $tc\phi^0$ to the flavor-mixing charged
coupling (FMCC) $bc\phi^\pm$ through the (s)quark-mixing matrix.
Therefore, if we can directly measure the coupling of FMCC at
future high energy colliders, it would provide the detailed
information on the FCNC and may give more precise constraints on
the FCNC than that inferred from kaon and bottom physics obtained
from low energy experiments.

\par
Although the electroweak corrections can contribute to the flavor
changing effect on the process \ppbchT, but the SUSY QCD
corrections via squark-gluino loops are dominant over the previous
ones, at least one order larger in magnitude. In this work, we
calculate the single charged Higgs boson production process
associated with a bottom-charm pair \ppbchT in the MSSM with
left-handed up-type squark mixings at QCD one-loop level at the
CERN LHC. We shall show the importance of squark-mixings in the
enhancement of production rate for \ppbchT process. We analyze the
SUSY QCD radiative contributions to the process \ppbchT by
adopting the relevant MSSM parameters at the Snowmass point SPS 4
with large $\tan\beta$. The paper is organized as below: In
section 2 we present a brief outline on the up-type squark mass
matrix considering the left-handed up-type squark mixings, and
diagonalize it to obtain the mass eigenstates matrix of squarks.
In section 3, we give the calculations of the cross sections of
\ppbch up to the order ${\cal O}(gg_s^4)$ in the MSSM. The
numerical results and discussions are presented in section 4.
Finally, a short summary is given.

\par
\section{Left-Handed Up-Type Squark Mixing}
\par
In the supersymmetric models, the SM flavor mixings between quarks
of three generations can be extended to include the superpartners
of quarks and leptons by introducing the supersymmetry
soft-breaking meachanism. These models leave further puzzles to
the flavor physics, since the soft-breaking Lagrangian of the
supersymmetry, which gives a mass spectrum of the supersymmetric
particles, involves numerous unconstrained free parameters. In
order to fit with low-energy FCNC data, we have to make specific
assumptions to these free parameters.

\par
In the study of the $bcH^{\pm}$ production at hadron colliders,
only the flavor mixing in squark sector is concerned. For the
down-type squark mixings, there are possible strong constraints on
the mixing parameters from the low-energy experimental data. For
example, the mechanism with down-type squark mixings in large
$\tan\beta$ could enhance the FCNC $B-$decays by several
orders\cite{Xiong} and seems to be ruled out by $B$-factory
experiments. But the up-type squark mixing between $\tilde t$ and
$\tilde c$ is subject to no strong low-energy
constraint\cite{Gabbiani}. Such $\tilde t$-$\tilde c$ squark
mixing is well motivated in low-energy supergravity models
(SUGRA)\cite{Duncan}. Ref.\cite{Hikasa} shows that at low energy
the $\tilde t$-$\tilde c$ mixing may be significant due to very
heavy top quark, and the mixing between $\tilde t_L$ and $\tilde
c_L$ is most likely to be large, which is proportional to a sum of
some soft masses. For theoretical simplicity in this work, we
focus on the MSSM with the squark-mixing assumption that only the
left-handed up-type squarks in three generations can mix with each
other\cite{sher}. In the super Cabibbo-Kobayashi-Maskawa(CKM)
basis $\tilde{U}^{\prime} = (\tilde u_L,\tilde u_R,\tilde
c_L,\tilde c_R,\tilde t_L,\tilde t_R)$, the $6\times6$ squark mass
matrix ${\cal M}^2_{\tilde{U}}$ of up-type squark sector takes the
form as\cite{matrix}
\begin{eqnarray}
 {\cal M}^2_{\tilde{U}}
      &=&
    \left(
    \begin{array}{cccccc}
      M^2_{L,u} &
      a_{u} m_{u} &
      \lambda_{12} M_{L,u} M_{L,c} &
      0 &
      \lambda_{31}^{\ast} M_{L,u} M_{L,t} &
      0 \\
      a^{\ast}_{u} m_{u} &
      M^2_{R,u} &
      0 &
      0 &
      0 &
      0 \\
      \lambda_{12}^{\ast} M_{L,c} M_{L,u} &
      0 &
      M^2_{L,c} &
      a_{c} m_{c} &
      \lambda_{23} M_{L,c} M_{L,t} &
      0 \\
      0 &
      0 &
      a^{\ast}_{c} m_{c} &
      M^2_{R,c} &
      0 &
      0 \\
      \lambda_{31} M_{L,t} M_{L,u} &
      0 &
      \lambda_{23}^{\ast} M_{L,t} M_{L,c} &
      0 &
      M^2_{L,t} &
      a_{t} m_{t} \\
      0 &
      0 &
      0 &
      0 &
      a^{\ast}_{t} m_{t} &
      M^2_{R,t}
      \end{array}
       \right),~~~
\end{eqnarray}
where
\begin{eqnarray}
\nonumber M^2_{L,q} &=& M^2_{\tilde{Q}, q} +
m^2_{Z}\cos2\beta(\frac{1}{2}-Q_q \sin^2 \theta_{W})+m^2_q, \\
\nonumber
M^2_{R,q} &=& M^2_{\tilde{U}, q} +  Q_q m^2_{Z}\cos2\beta \sin^2 \theta_{W}+m^2_q, \\
a_q &=& A_q   -  \mu \cot \beta, ~~~~~~~~~~~~ (q = u, ~ c, ~ t).
\end{eqnarray}
with $m_{Z}$ being the mass of $Z^0$, and $m_q$, $Q_q$ the
up-quark mass and charge. $M_{\tilde{Q}, q}$ and $M_{\tilde{U},
q}$ are mass parameters of supersymmetry soft breaking. $A_q(q =
u,c,t)$ are the trilinear scalar coupling parameters of Higgs
boson with two scalar quarks. $\mu$ is the mass parameter of the
Higgs boson sector and $\tan\beta$ is the ratio of the vacuum
expectation values in this sector. $\sin\theta_W$ contains the
electroweak mixing angle $\theta_W$. $\lambda_{12}$,
$\lambda_{23}$ and $\lambda_{31}$ are the flavor mixing strengths
of the $\tilde{u}_L$-$\tilde{c}_L$, $\tilde{c}_L$-$\tilde{t}_L$
and $\tilde{t}_L$-$\tilde{u}_L$ sectors, respectively. Since we
don't consider the CP-violation, all these squark-mixing
parameters have real and positive values varying in the range of
$[0, ~ 1]$.

\par
To obtain the mass eigenstates of the up-type squarks, we should
introduce an unitary matrix ${\cal R}^{(U)}$ defined as
\begin{eqnarray}
\tilde{U}^{\prime} = {\cal R}^{(U)} \tilde{U},
\end{eqnarray}
where
\begin{eqnarray}
\tilde{U}^{\prime}= \left(
    \begin{array}{c}
    \tilde{u}_{L} \\
    \tilde{u}_{R} \\
    \tilde{c}_{L} \\
    \tilde{c}_{R} \\
    \tilde{t}_{L} \\
    \tilde{t}_{R} \\
    \end{array}
\right),~~~~~~~~~ \tilde{U}= \left(
    \begin{array}{c}
    \tilde{u}_{1} \\
    \tilde{u}_{2} \\
    \tilde{u}_{3} \\
    \tilde{u}_{4} \\
    \tilde{u}_{5} \\
    \tilde{u}_{6}   \end{array}\right)
= \left(
    \begin{array}{c}
     \tilde{u}_{1} \\
    \tilde{u}_{2} \\
    \tilde{c}_{1} \\
    \tilde{c}_{2} \\
    \tilde{t}_{1} \\
    \tilde{t}_{2} \\
    \end{array}
\right).
\end{eqnarray}
The up-squark mass matrix $M_{\tilde{U}}^2$ is diagonalized by the
$6 \times 6$ matrix ${\cal R}^{(U)}$ via
\begin{eqnarray}
{\cal R}^{(U) \dag}{\cal M}^2_{\tilde{U}} {\cal R}^{(U)}= {\rm
diag} \{m^2_{\tilde{u}_1},...,m^2_{\tilde{u}_6}\}.
\end{eqnarray}
where $m^2_{\tilde{u}_j} ~ (j = 1,...,6)$ are the masses of mass
eigenstates of the six up-type squarks which depend on
$\lambda_{12}$, $\lambda_{23}$ and $\lambda_{31}$.

\par
\section{Calculation of the process \ppbchT }
\par
\begin{figure}[hbtp]
\vspace*{-1cm} \centerline{ \epsfxsize = 8cm \epsfysize = 3cm
\epsfbox{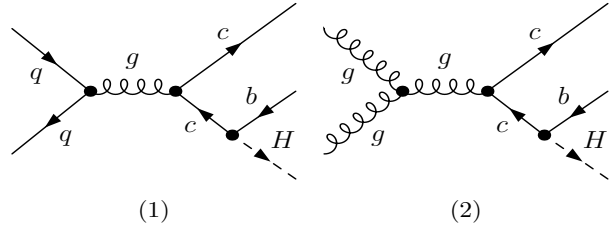}}  \vspace*{0cm}\caption{\em (1) One of the
tree-level (${\cal O}(gg_s^2 )$) Feynman diagrams for $q\bar q\to
\bar bcH^-$ subprocess. (2) One of the tree-level (${\cal
O}(gg_s^2)$) Feynman diagrams for $gg \to \bar bcH^-$ subprocess.
}\label{qgbch}
\end{figure}

The exclusive process of single charged Higgs boson production
associated with a bottom-charm quark pair, \ppbchT, involves the
contributions from the subprocesses of $q\bar q (q=u, d, c, s)$
annihilation and gluon-gluon fusion. Since the processes $q\bar
q/gg \to b\bar cH^+$ have the same total and differential cross
sections as their corresponding charge-conjugate subprocesses
$q\bar q/gg \to \bar bcH^-$ in the CP-conserving MSSM, we present
here only the calculations of the process $pp \to \bar bcH^-+X$.
For each subprocesses of $q\bar q \to \bar bcH^-$ and $gg \to \bar
bcH^-$, we depict one tree-level(${\cal O}(gg_s^2)$) Feynman
diagram as a demonstration in Fig.\ref{qgbch}(1) and
Fig.\ref{qgbch}(2), respectively. \vspace*{1.0cm}

The tree-level total cross section of \ppbch can be obtained by
doing following integration:
\begin{eqnarray}\nb
\sigma_{tree}(pp(AB) \to \bar bcH^- + X)=~~~~~~~~~~~~~~~~~~~~~~~~~~~~~~~~~~~~~~~~~~~~~~~~~~~~~~~~\\
\sum^{s\bar s,c\bar c,gg}_{ij=u\bar u,d\bar
d}\frac{1}{1+\delta_{ij}}\int dx_Adx_B
[G_{i/A}(x_A,\mu_f)G_{j/B}(x_B,\mu_f)\hat
\sigma^{ij}_{tree}(x_A,x_B,\mu_f)+(A \leftrightarrow B)],~~~
\label{PDFint}
\end{eqnarray}
where $x_A$ and $x_B$ are defined as
\begin{equation}
x_A=\frac{p_1}{P_A}, x_B=\frac{p_2}{P_B}, \label{PDFint1}
\end{equation}
where A and B represent the incoming colliding protons. $p_1$,
$p_2$, $P_A$ and $P_B$ are the momenta of partons and protons.
$\hat \sigma ^{ij}_{tree}(ij=u\bar u, d\bar d, c\bar c, s\bar s,
gg)$ is the total LO cross section at parton-level for incoming i
and j partons. $G_{i/A(B)}$ is the leading-order parton
distribution function (PDF) for parton i in hadron A(B). We adopt
CTEQ6L1 PDF in the calculation of the tree-level cross
section\cite{CTEQ6M}.

\par
In the calculation of the SUSY QCD NLO contributions in the
framework of the MSSM with left-handed up-type squark-mixings, we
adopt the 'tHooft-Feynman gauge, and use dimensional
regularization(DR) method in $D=4-2\epsilon$ dimensions to isolate
the ultraviolet(UV), soft and collinear infrared(IR)
singularities. The modified minimal subtraction ($\overline{\rm
MS}$) scheme is employed to renormalize and eliminate UV
divergency. The SUSY QCD NLO contributions can be divided into two
parts: the virtual contributions from one-loop diagrams, and the
real gluon/light-quark emission contributions.

\par
\begin{figure}[hbtp]
\vspace*{-1cm} \centerline{ \epsfxsize = 8cm \epsfysize = 3cm
\epsfbox{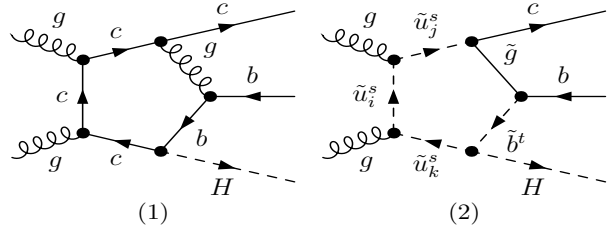}}  \vspace*{0cm}\caption{\em The representative
QCD pentagon Feynman diagrams for $gg\to \bar bcH^-$ subprocess.
(1) is the SM-like diagram, and (2) is the SUSY QCD one-loop
diagram. The lower-index i in $\tilde{q}_i^s$ implies the $i$-th
generation$(i=1,2,3)$ and the upper-index $s=1,2$.} \label{loop}
\end{figure}

The unrenormalized virtual contribution to the subprocess \qgbch
in the MSSM consists of self-energy, vertex, box, and pentagon
diagrams. These one-loop diagrams for subprocess \qgbch can be
divided into two parts: One is the SM-like part which comprises
the diagrams including gluon/quark loops. Another is called SUSY
part involving virtual gluino/squark exchange loops. In the later
part, contributions from the one-loop diagrams with left-handed
up-type squark mixings between different generations are
considered. For demonstration, we plot the QCD one-loop pentagon
diagrams for the $gg\to b\bar cH^+$ subprocess in Fig.\ref{loop}.
The figures in Fig.\ref{loop}(1)-(12) belong to the the SM-like
part, while Fig.\ref{loop}(13)-(24) to the SUSY part. The
amplitude for the virtual SM-like contribution part contains both
ultraviolet(UV) and soft/collinear infrared(IR) singularities,
while the amplitude corresponding to SUSY loop part contains only
UV singularities. In order to remove the UV divergences, we
renormalize the relevant fields, the masses of charm- and
bottom-quark in propagators and the $bcH^\pm$ Yukawa coupling by
adopting on-shell(OS) scheme. The renormalization constants of the
CKM-matrix elements $V_{ij}(i,j=1,2,3)$ can be obtained by keeping
the unitarity of the renormalized CKM-matrix, and expressed
as\cite{Denner}
\begin{equation}
\delta V_{ij}=\frac{1}{4}\left[\left(\delta Z_{ik}^{u,L}-\delta
Z_{ik}^{u,L\dagger}\right)V_{kj}- V_{ik}\left(\delta
Z_{kj}^{d,L}-\delta Z_{kj}^{d,L,\dagger}\right)\right].
\end{equation}

\vspace*{1.5cm}

\par
We use the $\overline{MS}$ mass of the bottom
quark($\overline{m}_b (\mu_r)$) in $bcH^\pm$ Yukawa coupling, to
absorb the large logarithms contributions which arise from the
renormalization of bottom quark mass\cite{mb}, but keep the bottom
quark pole mass everywhere else. The bottom quark mass in
propagator is renormalized by adopting the on-shell(OS) scheme.
The expressions of the $\overline{MS}$ mass of the bottom quark
$\overline{m}_b(\mu_r)$ corresponding 1-loop and 2-loop
renoramlization groups are given by
\begin{eqnarray}\label{mboneloop}
\overline{m}_b(\mu_r)_{1-loop}&=&m_b\left[\frac{\alpha_s(\mu_r)}{\alpha_s(m_b)}\right]^{c_0/b_0},
\end{eqnarray}
\begin{eqnarray}\nb
\overline{m}_b(\mu_r)_{2-loop}&=&m_b\left[\frac{\alpha_s(\mu_r)}{\alpha_s(m_b)}\right]^{c_0/b_0}
\left[1+\frac{c_0}{b_0}(c_1-b_1)\frac{\alpha_s(\mu_r)-\alpha_s(m_b)}{\pi}\right]\left(1-\frac{4}{3}\frac{\alpha_s(m_b)}{\pi}\right),
\\ \label{mbtwoloop}
\end{eqnarray}
where
\begin{eqnarray}\nb
b_0&=&\frac{1}{4\pi}\left(\frac{11}{3}N_c-\frac{2}{3}n_{f}\right),~~~c_0=\frac{1}{\pi},\\
b_1&=&\frac{1}{2\pi}
\frac{51N_c-19n_{f}}{11N_c-2n_{f}},~~~~c_1=\frac{1}{72\pi}\left(101N_c-10n_{f}\right)
\end{eqnarray}
In our calculation we adopt $\overline{m}_b(\mu_r)_{1-loop}$ in
Eq.(\ref{mboneloop}) and $\overline{m}_b(\mu_r)_{2-loop}$ in
Eq.(\ref{mbtwoloop}) as the $\overline{m}_b(\mu_r)$ mass for the
LO and NLO cross sections, respectively\cite{Dawson}. The
renormalization of the bottom quark mass in Yukawa coupling is
defined as
\begin{equation}
m^0_b=\overline{m}_b(\mu_r)
\left[1+\delta^{SM-like}+\delta^{SUSY}\right],
\end{equation}
where the counterterm of the SM-like QCD part $\delta^{SM-like}$
is calculated in $\overline{MS}$ scheme, while SUSY counterterm
part $\delta^{SUSY}$ is calculated in on-shell(OS) scheme. Due to
the fact that there are significant corrections to $bcH^\pm$
coupling for large value of $\tan\beta$, we absorb these
corrections in the Yukawa coupling\cite{dmb}. The resumed
$bcH^\pm$ Yukawa coupling can be expressed as\cite{wp,mb,Guasch}
\begin{equation}
g_{\bar bcH^-} = \frac{igV_{cb}}{\sqrt{2}m_W}\left \{m_c \cot\beta
P_R+\overline{m}_b(\mu_r)\frac{1-\frac{\Delta_b}{\tan^2\beta}}
{1-\Delta_b}\tan\beta P_L\right \},
\end{equation}
where
\begin{eqnarray}\nb
\Delta_b&=&\frac{\Delta m_b}{1+\Delta_1},\\\nb \Delta
m_b&=&\frac{2}{3} \frac{\alpha_s}{\pi} m_{\widetilde{g}}\mu
\tan\beta I(m^2_{\widetilde{b_3}}, m^2_{\widetilde{b_2}},
m^2_{\widetilde{g}}),\\\nb \Delta_1&=&-\frac{2}{3}
\frac{\alpha_s}{\pi} m_{\widetilde{g}}A_b I(m^2_{\widetilde{b}_1},
m^2_{\widetilde{b}_2},
m^2_{\widetilde{g}}),\\
I(a,b,c)&=&-\frac{ab \log\frac{a}{b}+bc \log{b}{c}+ca
\log\frac{c}{a}}{(a-b)(b-c)(c-a)}.
\end{eqnarray}

\par
Since in the calculation of the cross section of the process
\ppbch at the tree-level, we used the resumed $bcH^\pm$ Yukawa
coupling, we have to add a finite renormalization of the bottom
quark mass in $bcH^\pm$ Yukawa coupling to avoid double counting
in the SUSY QCD NLO corrections to the cross
section\cite{doublecouting}.
\begin{eqnarray}
m_b&\to& \overline{m}_b
(\mu_r)[1+\Delta^{H^-}_b]+O(\alpha_s^2),\\
\Delta_b^{H^-}&=&\frac{2}{3}\frac{\alpha_s}{\pi}(1+\frac{1}{\tan^2\beta})m_{\widetilde{g}}\mu
\tan\beta I(m^2_{\widetilde{b}_1}, m^2_{\widetilde{b}_2},
m^2_{\widetilde{g}}).
\end{eqnarray}

\par
For the renormalization of the strong coupling constant $g_s$, we
divide the counterterm of the strong coupling constant into two
terms: SM-like QCD term and SUSY term ($\delta g_s=\delta
g_s^{(SM-like)}+\delta g_s^{(SUSY)}$), and the explicit
expressions of these two terms can be obtained by adopting
$\overline{MS}$ scheme at renormalization scale
$\mu_r$\cite{wp,Reina}.
\begin{equation} \frac{\delta
g_s^{(SM-like)}}{g_s}=-\frac{\alpha_s (\mu_r)}{4\pi}
\left[\frac{\beta_0^{(SM-like)}}{2} \frac{1}{\bar \epsilon} +
\frac{1}{3} \ln \frac{m^2_t}{\mu^2_r}\right],
\end{equation}
\begin{equation}
\frac{\delta
g_s^{(SUSY)}}{g_s}=-\frac{\alpha_s(\mu_r)}{4\pi}\left[\frac{\beta_0^{(SQCD)}}{2}
\frac{1}{\bar \epsilon} + \frac{N_c}{3} \ln
\frac{m^2_{\widetilde{g}}}{\mu^2_r} + \sum^{i=1,2}_{U=u,c,t}
\frac{1}{12}\ln
\frac{m^2_{\widetilde{U}_i}}{\mu^2_r}+\sum^{j=1,2}_{D=d,s,b}
\frac{1}{12}\ln \frac{m^2_{\widetilde{D}_j}}{\mu^2_r}\right],
\end{equation}
where we have used the notations
\begin{equation}
\beta^{(SM-like)}_0=\frac{11}{3}N_c-\frac{2}{3}n_{f}-\frac{2}{3},~~~~
\beta^{(SUSY)}_0=-\frac{2}{3}N_c-\frac{1}{3}\left(n_{f}+1\right).
\end{equation}

\par
The number of colors $N_c$ equals 3, the number of active flavors
is taken to be $n_{f}=5$ and $1/\bar
\epsilon=1/\epsilon_{UV}-\gamma_E+\ln(4\pi)$. The summation is
taken over the indexes of squarks and generations. Since the
$\overline{MS}$ scheme violates supersymmetry, it is necessary
that the $q\tilde{q}\tilde{g}$ Yukawa coupling $\hat g_s$, which
should be the same with the $qqg$ gauge coupling $g_s$ in the
supersymmetry, takes a finite shift at one-loop order as shown in
Eq.(\ref{gs})\cite{gs}.
\begin{equation}
\hat{g}_s=g_s \left[1+
\frac{\alpha_s}{8\pi}\left(\frac{4}{3}N_c-C_F\right)\right],
\label{gs}
\end{equation}
with $C_F=4/3$. In our numerical calculation we take this shift
between $\hat g_s$ and $g_s$ into account.

\par
The SUSY QCD one-loop Feynman diagrams for both $q\bar q \to \bar
bcH^-$ and $gg \to \bar bcH^-$ subprocesses can be divided into
virtual gluon/quark exchange part(SM-like) and virtual
gluino/squark exchange part(SUSY). We express the renormalized
amplitudes for both subprocesses as
\begin{equation}
M^{q\bar q,gg}_{virtual}=M^{q\bar q,gg}_{SM-like}+M^{q\bar
q,gg}_{SUSY}
\end{equation}
Then the SUSY QCD NLO contributions to the cross sections of the
subprocesses \qgbch can be expressed as
\begin{equation}
\hat\sigma^{q\bar q,gg}_{virtual}=\int d\Phi_3
\overline{\sum}(2Re(M^{q\bar q,gg}_{tree}M^{q\bar
q,gg\dagger}_{virtual})+|M^{q\bar q,gg}_{tree}|^2),
\label{amplitudeSum}
\end{equation}
where $d\Phi_3$ is three-body phase space element, $M^{q\bar
q}_{tree}$ and $M^{gg}_{tree}$ are the Born amplitudes for \qgbch
subprocesses separately, and $M^{q\bar q}_{virtual}$ and
$M^{gg}_{virtual}$ are their renormalized amplitudes of the SUSY
QCD one-loop diagrams. The bar over the summation in
Eq.(\ref{amplitudeSum}) recalls averaging over initial spin and
color states.

\par
$\hat\sigma^{q\bar q,gg}_{virtual}$ are free of UV divergences but
contain soft/collinear IR divergences, among them the soft IR
divergence can be cancelled by adding with the soft real gluon
emission corrections. The soft and collinear singularities from
real gluon emission subprocess can be conveniently isolated by
slicing the phase space into different regions defined with
suitable cutoffs\cite{PSS}. We introduce an arbitrary soft cutoff
$\delta_s(\equiv 2E_{g}/\sqrt{\hat{s}})$ with small value to
separate the phase space of the gluon emission $2 \to 4$
subprocess into two regions, i.e., soft and hard gluon emission
regions. Then for the gluon emission $2 \to 4$ subprocesses based
on $(q\bar q,q'\bar {q'},gg) \to \bar bcH^-$, $(q=u,d,~q'=s,c)$,
we have
\begin{eqnarray}
\hat{\sigma}_{real}((q\bar{q},q'\bar{q'},gg) \to \bar
bcH^-g)&=&\hat{\sigma}_{soft}((q\bar{q},q'\bar{q'},gg) \to \bar
bc H^-g)  \nonumber \\
&+&\hat{\sigma}_{hard}((q\bar{q},q'\bar{q'},gg) \to \bar bc H^-g),
\end{eqnarray}
where $\hat{\sigma}_{soft}$ is obtained by integrating over the
soft region of the emitted gluon phase space, and contains all the
soft IR singularities. Furthermore, we decompose each cross
section of $\hat{\sigma}_{hard}$ for hard-gluon emission
subprocesses $q\bar{q}/q'\bar{q'}/gg \to \bar bc H^-g$ and
light-quark emission subprocesses $\hat{\sigma}_{real}$ for $(\bar
qg,qg) \to \bar bc H^-(\bar q,q)$, into a sum of collinear and
non-collinear terms to isolate the remaining collinear
singularities from $\hat{\sigma}_{hard}$ and $\hat{\sigma}_{real}$
, by introducing another cutoff $\delta_c$ called collinear
cutoff.

\par
The cross sections in the non-collinear hard-gluon/light-quark
emission regions, can be obtained by performing the phase space
integration in 4-dimension by using Monte Carlo method. Then we
get the SUSY QCD NLO corrected cross sections for subprocesses
$q\bar q/q'\bar {q'}/gg \to \bar bcH^-$ as follows. For $q\bar q$
annihilation($q=u,d$) subprocesses,
\begin{eqnarray}
\label{subCross1} \hat{\sigma}_{loop}(q\bar q \to \bar bc
H^-)&=&\hat{\sigma}_{tree}(q\bar q \to \bar bc H^-)+
\hat{\sigma}_{virtual}(q\bar q \to \bar bc H^-)   \nonumber \\
&+&\sum_{ij=q\bar q,}^{\bar qg,qg}\hat{\sigma}_{real}(ij \to \bar
bc H^-(g,\bar q,q)),
\end{eqnarray}
while for $q'\bar {q'}$ annihilation($q'=c,s$) and $gg$ fusion
subprocesses,
\begin{eqnarray}
\label{subCross2} \hat{\sigma}_{loop}(q'\bar {q'}/gg \to \bar bc
H^-)&=&\hat{\sigma}_{tree}(q'\bar {q'}/gg \to \bar bc H^-)+
\hat{\sigma}_{virtual}(q'\bar {q'}/gg \to \bar bc H^-)   \nonumber \\
&+&\hat{\sigma}_{real}(q'\bar {q'}/gg \to \bar bc H^-g),
\end{eqnarray}
The remaining collinear IR divergences in Eq.(\ref{subCross1}) and
Eq.(\ref{subCross2}) are absorbed into the parton distribution
functions, separately. Then the SUSY QCD NLO corrected total cross
section for $pp \to \bar bcH^- +X$ process $\sigma_{loop}$, can be
obtained by using Eq.(\ref{PDFint}) and replacing
$\hat{\sigma}^{ij}_{tree}(ij \to \bar bc H^-)$, $(ij=q\bar
q,q'\bar {q'},gg)$ by $\hat{\sigma}_{loop}(ij \to \bar bc H^-)$,
$(ij=q\bar q,q'\bar {q'},gg,\bar qg,qg)$, and CTEQ6L1 parton
distribution functions by CTEQ6M ones\cite{CTEQ6M}. The total
cross section $\sigma_{loop}(pp \to \bar bcH^- +X)$ up to SUSY QCD
NLO should be independent on the two arbitrary cutoffs $\delta_s$
and $\delta_c$. In our both analytical and numerical calculations,
we checked the cancellations of the UV and IR divergences, and
found that the final results are both UV- and IR-finite.

\par
\section{Numerical Results and Discussions}
\par
In this section, we present some numerical results of the total
and differential cross sections of the precess \ppbchT inspired by
the squark-mixing loop contributions at the LHC. We take the SM
parameters as: $m_W=80.425~GeV$, $m_Z=91.1876~GeV$, $m_t=175~GeV$,
$m_b=4.7~GeV$, $m_c=1.2~GeV$, $m_s=0.15~GeV$, $V_{cb}=0.04$,
$V_{ub}=0.0035$, $V_{cd}=0.222$, $V_{cs}=0.97415$, $V_{cb}=0.04$
and $V_{tb}=0.99915$ \cite{parameter} and neglect the light-quark
masses$(m_{u,d})$ in the numerical calculation. The value of the
fine structure constant at the energy scale of $Z^0$ pole mass, is
taken as $\alpha_{ew}(m_Z)^{-1}=127.918$\cite{parameter}. We use
the CTEQ6L1 and CTEQ6M parton distribution functions for the
calculations of LO and NLO contributed cross sections,
respectively \cite{CTEQ6M}. The colliding energy of proton-proton
collider at the LHC is $\sqrt{s}=14~TeV$. We fix the value of the
renormalization/factorization scale being $Q=Q_0=\mu_r=\mu_f$ for
simplicity, where $Q_0$ is defined to be a half of the final
particle masses. As a numerical demonstration, we refer to the
relevant MSSM parameters of Snowmass point SPS 4 with high
$\tan\beta$\cite{SPS4,SPS4-1} except considering the left-handed
up-type squark mixings in the MSSM. SPS 4 is a mSUGRA point with
input parameters:
\begin{equation}
m_0=400~GeV,~~m_{1/2}=300~GeV,~~A_0=0,~~\tan\beta=50,~~\mu > 0.
\label{SPS 4}
\end{equation}
Since the squark sector in Snowmass points doesn't include squark
mixing, we don't adopt the SPS 4 physical sparticle spectrum, but
the ISAJET\cite{ISAJET} equivalent input MSSM parameters at this
benchmark point, with which one can reproduce the ISAJET spectrum
with SUSYGEN\cite{SUSYGEN} and PYTHIA\cite{PYTHIA}. These relevant
MSSM parameters at SPS 4 point we used in our calculation, are
listed below\cite{SPS4-1}.
\begin{eqnarray}\nb
\tan\beta=50,~~m_{H^\pm}=416.28GeV,~~  m_{\tilde
g}=721.03GeV,\\\nb m_{\tilde u_L}=m_{\tilde d_L}=m_{\tilde
c_L}=m_{\tilde s_L}=732.2GeV,~~ m_{\tilde b_L}=m_{\tilde
t_L}=640.09GeV,\\\nb m_{\tilde u_R}=m_{\tilde c_R}=716.00GeV, ~~
~~ m_{\tilde d_R}=m_{\tilde s_R}=713.87GeV,~~  m_{\tilde
b_R}=673.40GeV,\\ m_{\tilde t_R}=556.76GeV,~~  A_t=-552.20GeV,
~~ A_b=-729.52GeV.\label{parameter}
\end{eqnarray}

\par
The squark-mixing parameters $\lambda_{12}$, $\lambda_{31}$ and
$\lambda_{23}$ are constrained by low energy data on
FCNC\cite{Gabbiani}. Following the reference \cite{Gabbiani}, we
use the bounds for the squark-mixing parameters as
\begin{eqnarray}\nb
\lambda_{12}&<& 0.1 \sqrt{m_{\widetilde u} m_{\widetilde
c}}/500~GeV,   \\\nb
\lambda_{31}&<& 0.098 \sqrt{m_{\widetilde u} m_{\widetilde t}}/500~GeV,   \\
\lambda_{23}&<& 8.2 m_{\widetilde c} m_{\widetilde t}/(500~GeV)^2.
\label{constraints}
\end{eqnarray}

Considering above limitations on squark-mixing parameters, in our
calculations we use the MSSM parameters shown in
Eq.(\ref{parameter}) and take $\lambda_{12} = 0.03$, $\lambda_{31}
= 0.03$ and $\lambda_{23} = \lambda = 0.6$, which satisfy the
constraints of Eq.(\ref{constraints}), if there is no different
statement. Actually, our calculation shows that the cross section
involving NLO QCD corrections for process \ppbchT is mostly
related to the squark-mixing parameter $\lambda_{23}$, but not
sensitive to the $\lambda_{12}$ and $\lambda_{31}$. That implies
the contributions from the $\tilde{u}_L$-$\tilde{c}_L$ and
$\tilde{u}_L$-$\tilde{t}_L$ squark-mixings are very small in our
chosen parameter space.

\par
In numerical calculation, we put the cuts on the transverse
momenta of final bottom, charm quarks and charged Higgs boson as
\begin{equation}
|p^b_T|>20~GeV,~~~ |p^c_T|>20~GeV,~~~|p^{H^-}_T|>10~GeV.
\end{equation}

\par
As we discussed in above section, the final results should be
independent on cutoffs $\delta_s$ and $\delta_c$. That is also one
of the checks of the correctness of our calculation. We checked
the independence of cutoffs $\delta_s$ and $\delta_c$ in our
calculation. In the following numerical calculation we fix
$\delta_s=10^{-3}$ and $\delta_c=10^{-5}$.
\begin{figure}[hbtp] \centerline{
\epsfxsize = 8cm \epsfysize = 7cm \epsfbox{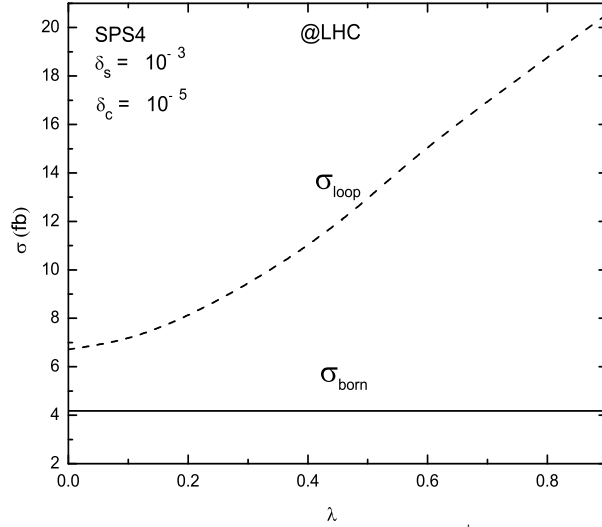}}
\vspace*{-0.5cm}\caption{\em The cross sections for the process
\ppbchT at the tree-level(${\cal O}(gg_s^2)$) $\sigma_{tree}$ and
up to QCD one-loop level(${\cal O}(gg_s^4)$) $\sigma_{loop}$ as
the functions of the mixing strength parameter
$\lambda(=\lambda_{23})$ in the up-type squark mass matrix at the
LHC. The relevant MSSM parameters at the Snowmass point SPS 4 are
adopted(shown in Eq.(\ref{parameter})).} \label{lambda}
\end{figure}

\par
In Fig.\ref{lambda} we present the total cross sections for the
process \ppbchT at the tree-level($\sigma_{tree}$) and up to SUSY
QCD NLO($\sigma_{loop}$) as the functions of the mixing strength
between $\tilde{c}_L$ and $\tilde{t}_L$, $\lambda(=\lambda_{23})$,
at the LHC with the relevant MSSM parameters at the Snowmass point
SPS 4 shown in Eq.(\ref{parameter}). We can see the one-loop
correction significantly enhances the corresponding LO cross
section(solid curve), and loop contribution goes up with the
increment of the mixing parameter $\lambda$. As we know the
tree-level cross section for \ppbchT process is suppressed by
CKM-matrix element in the Yukawa coupling of $bcH^\pm$. But the
mixing between $\tilde{c}_L$ and $\tilde{t}_L$ in the SUSY
soft-breaking sector can make the one-loop contributions quite
sizable, and significantly increases the tree-level cross section
by a factor of $1.5 \sim 5$, which is shown in Fig.\ref{lambda}.
\begin{figure}[hbtp] \centerline{
\epsfxsize = 6.8cm \epsfysize = 7cm
\epsfbox{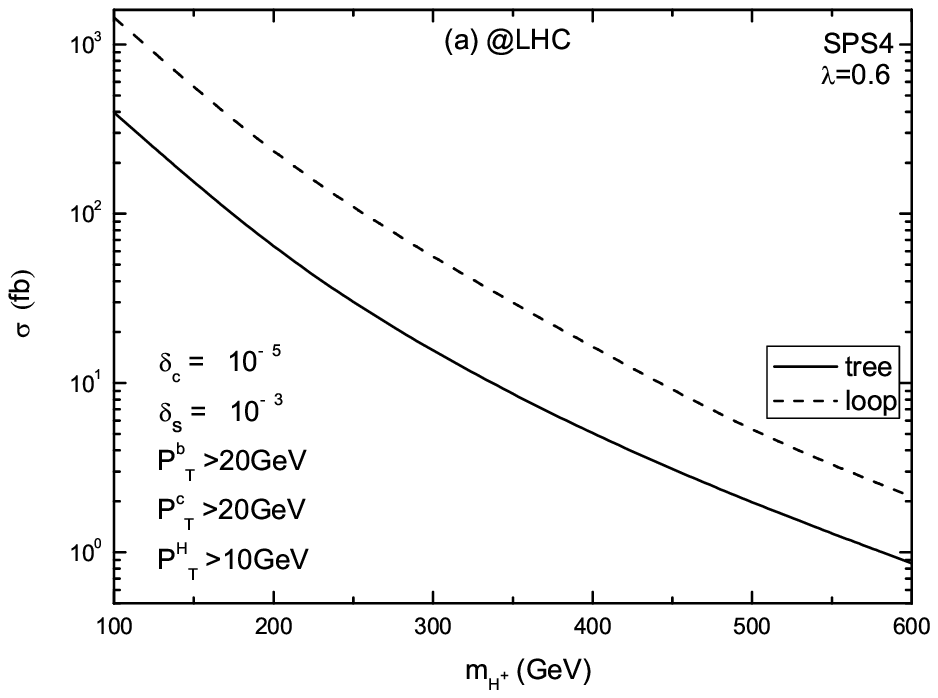}\epsfxsize = 6.8cm \epsfysize = 7cm
\epsfbox{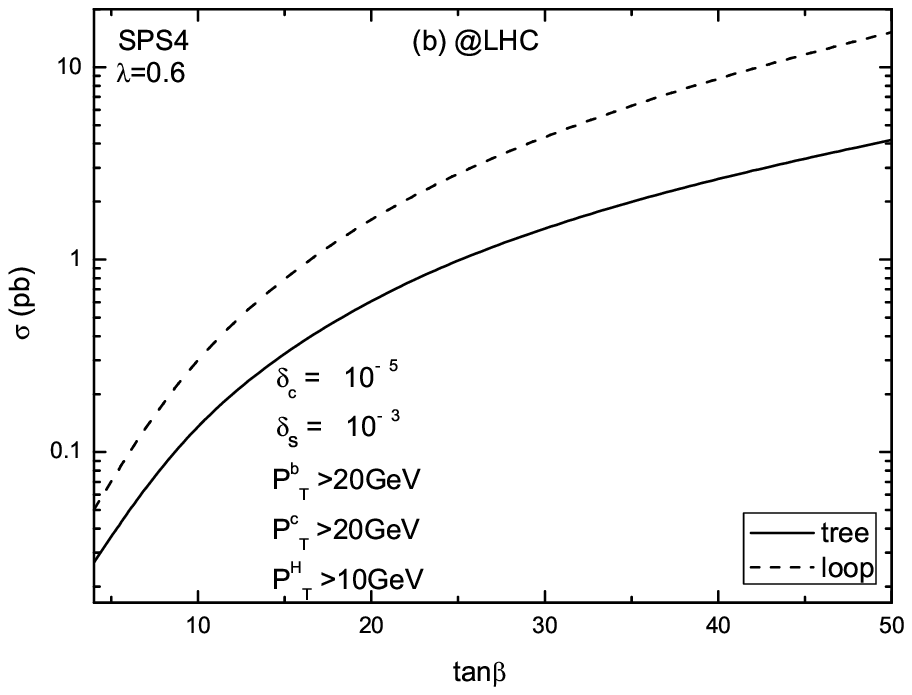}} \vspace*{-0.5cm}\caption{\em The cross
sections for the process \ppbchT at the tree-level
$\sigma_{tree}$ and up to SUSY QCD NLO $\sigma_{loop}$ as the functions
of the mass of charged Higgs boson(Fig.4(a)) and the ratio of the
vacuum expectation values $\tan\beta$(Fig.4(b)). The other relevant
MSSM parameters are taken from the Snowmass point SPS 4 listed
in Eq.(\ref{parameter}).}\label{mhtb}
\end{figure}

\par
In Fig.\ref{mhtb} we depict the cross sections for the process
\ppbchT at the tree-level and up to SUSY QCD NLO at the LHC as the
functions of the mass of charged Higgs boson in Fig.\ref{mhtb}(a)
and the ratio of the vacuum expectation values $\tan\beta$ in
Fig.4(b) respectively, by taking the other MSSM parameter values
from the Snowmass point SPS 4(see Eq.(\ref{parameter})). We can
read from Fig.\ref{mhtb}(a) that when $m_{H^{\pm}}$ increases from
$100~GeV$ to $600~GeV$, the total NLO QCD cross section decreases
from $1.548~pb$ to $2.13~fb$ at the LHC, while the tree-level
cross section decreases from $0.40~pb$ to $0.86~fb$. We can see
that if the MSSM scenario with left-handed up-type squark-mixings
is really true, the LHC machine with an integrated luminosity of
about $100~fb^{-1}$ can have the potential to find the signature
of charged Higgs boson via the process \ppbchT with $m_{H^\pm}$ in
the range of $[100~GeV,~600~GeV]$. Fig.4(b) shows that with
$m_{H^\pm}=416.28~GeV$ the cross section up to the ${\cal
O}(gg_s^4)$ order $\sigma_{loop}$ increases from $0.05 ~fb$ to
$15.18~fb$, as $\tan\beta$ varies from 4 to 50, while the
tree-level cross section is much smaller than $\sigma_{loop}$. We
can conclude that the production rate of the single charged Higgs
bosons associated with bottom-charm quark pair at the LHC, can be
enhanced by the gluino/squark loop contributions in the MSSM with
squark-mixing structure.
\begin{figure}[hbtp] \centerline{
\epsfxsize = 6.8cm \epsfysize = 7cm \epsfbox{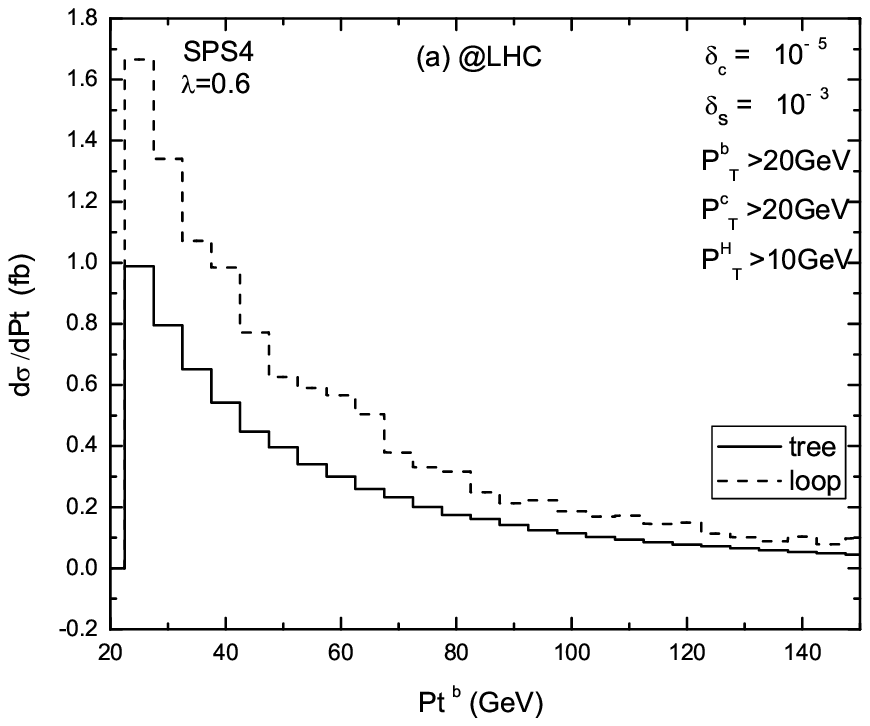}\epsfxsize
= 6.8cm \epsfysize = 7cm \epsfbox{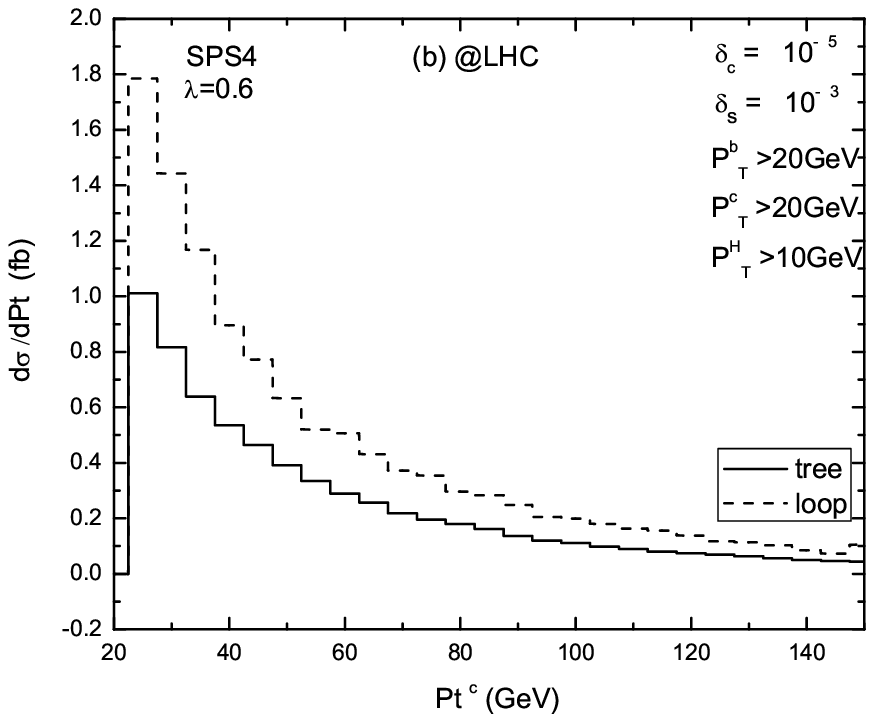}\epsfxsize = 6.8cm
\epsfysize = 7cm \epsfbox{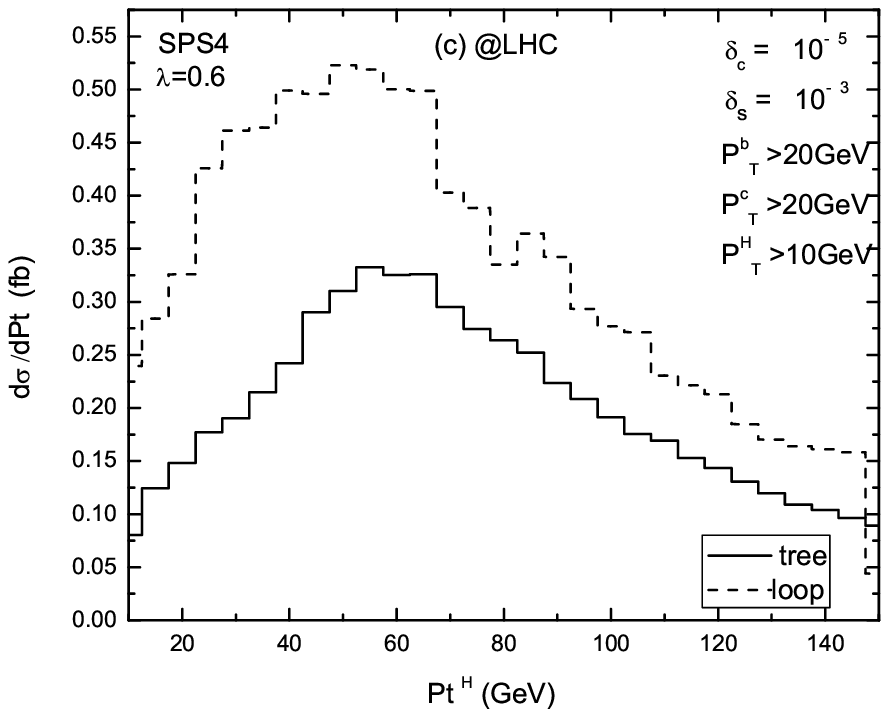}} \vspace*{-0.5cm}\caption{\em
The differential cross sections of the \ppbchT process at the
tree-level(${\cal O}(gg_s^2)$) and up to the SUSY QCD NLO(${\cal
O}(gg_s^4)$) as the functions of (a) the transverse momentum of
bottom-quark $p_T^{b}$, (b) the transverse momentum of charm-quark
$p_T^{c}$, and (c) the transverse momentum of the charged Higgs
boson $p_T^{H^\pm}$ at the LHC, with the relevant parameters of
the Snowmass point SPS 4(shown in
Eq.(\ref{parameter}).}\label{pt-bch}
\end{figure}

\par
In Figs.\ref{pt-bch}(a-c) we present the distributions of the
differential cross sections $d\sigma_{tree,loop}/dp_T^{b}$,
$d\sigma_{tree,loop}/dp_T^{c}$, $d\sigma_{tree,loop}/dp_T^{H^\pm}$
for the process \ppbchT in the MSSM with left-handed up-type
squark-mixing structure and the relevant parameters at the
Snowmass point SPS 4(shown in Eq.(\ref{parameter})) at the LHC.
These figures demonstrate that the loop contributions up to the
order (${\cal O}(gg_s^4)$) can significantly enhance the
tree-level differential cross sections $d\sigma_{tree}/dp_T^{b}$,
$d\sigma_{tree}/dp_T^{c}$ and $d\sigma_{tree}/dp_T^{H^\pm}$ at the
LHC. We find that in the low $p_T^b$ and $p_T^c$ regions the
corresponding differential cross section values including SUSY QCD
NLO corrections can be very large.

\par
\section{Summary}
\par
The general three-family up-type squark mass matrix originating
from the soft SUSY breaking sector, can induce the cross section
enhancement for the \ppbchT process at one-loop level. In this
paper we investigate the flavor changing effects on the production
of single charged Higgs bosons in association with $b$-$c$ quark
pair in the framework of the MSSM with left-handed up-type squark
mixings. We analyze the dependence of the cross section involving
NLO QCD corrections for \ppbchT process on the charged Higgs-boson
mass $m_{H^{\pm}}$, the ratio of the vacuum expectation values
$\tan\beta$, and the distributions of the transverse momenta of
bottom-quark $p_T^b$, charm-quark $p_T^c$ and charged Higgs boson
$p_T^{H^\pm}$ at the CERN LHC. We find that the one-loop
contributions are mostly inspired by the mixing of the
$\tilde{c}_L$-$\tilde{t}_L$ sector, and the QCD NLO corrections
can significantly enhance the corresponding tree-level(${\cal O}(g
g_s^2)$) cross sections in the MSSM with squark-mixings. Our
numerical results show the corrected cross section can reach the
value of $1.548~pb$ in our chosen parameter space. With this
production rate we may discriminate models of flavor symmetry
breaking and reveal new exciting discovery channel for the
signature of single charged Higgs boson at the LHC.

\vskip 5mm
\par
\noindent{\large\bf Acknowledgments:} This work was supported in
part by the National Natural Science Foundation of China, the
Education Ministry of China and a special fund sponsored by
Chinese Academy of Sciences.

\end{document}